\documentclass[prl,twocolumn,aps,amssymb,showpacs,floatfix,groupedaddress]{revtex4}
\usepackage{bm}
\usepackage{amssymb}
\usepackage{times}
\usepackage{latexsym}
\usepackage{graphicx}
\usepackage{color}
\usepackage{subfigure}
\usepackage{eqnarray}
\usepackage{amsmath}

\begin{document}
\title{Analytical calculation of electron's group velocity surfaces in uniform strained graphene}
\author{ Wilfrido A. G\'{o}mez-Arias$^1$ and Gerardo G. Naumis$^{1,2}$}
 
\affiliation{1. Departamento de F\'{i}sica-Qu\'{i}mica, Instituto de
F\'{i}sica, Universidad Nacional Aut\'{o}noma de M\'{e}xico (UNAM),
Apartado Postal 20-364, 01000 M\'{e}xico, Distrito Federal,
M\'{e}xico} 
\affiliation{2. Physics Department, George Mason University, Fairfax, Virginia, USA} 
 

\begin{abstract}
Electron group velocity for graphene under uniform strain is obtained analitically by using 
the Tight-Binding approximation.  Such closed analytical expressions are useful in order to 
calculate electronic, thermal and optical properties of strained graphene.  
These results allow to understand 
the behavior of electrons when graphene is subjected to strong strain and nonlinear corrections, for which 
the usual Dirac approach  is not longer valid. Some particular cases 
of uniaxial and shear strain were analized. The evolution of the electron group velocity indicates a break up of the
trigonal warping symmetry, which is replaced by a warping consistent with the symmetry of the strained
reciprocal lattice. The Fermi velocity becomes strongly anisotropic, i.e, for a strong pure shear-strain (20\% of the lattice parameter), the 
two inequivalent Dirac cones merge and the Fermi velocity
is zero in one of the principal axis of deformation. We found that non-linear terms
are essential to describe the effects of deformation for electrons near or at the Fermi energy.  
\end{abstract}
\maketitle

\section{Introduction}
Graphene was the first two-dimensional (2D) crystal discovered \cite{Novoselov04}. It has 
been broadly studied due to the observed peculiar physical 
properties \cite{Geim07,Sarma07, Geim09,Novoselov11}. The electronic
properties are mainly determined by electrons at the Fermi energy \cite{Geim09}. For graphene, 
such electrons have momentum near or at the high symmetry points of Brillouin Zone (BZ).
This behavoir can be modeled by a Dirac Hamiltonian \cite{Neto09}, 
where electrons behave as massless Dirac fermions with a Fermi velocity $v_f\simeq1\times10^6$ m/s, 
which plays the role of the speed of light. In this model, the Fermi velocity is a constant parameter. 
However, this is not longer true when graphene has corrugations (curved space) or is stretched, since 
these deformations give rise to a space-dependent Fermi velocity \cite{Pereira09a}, suggesting changes 
in the electronic conductivity.  Furthermore, in the case of stretching, a bandgap opening is 
observed \cite{Pereira09a,Colombo10,Sena12}. Such results open the possibility for doing 
``strain engineering'' in order to tailor the electronic properties and thus control the electron 
transport \cite{Pereira09b,Guinea12,Guinea13,Falko13,SalvadorSSC,Nguyen14}. 

Several theoretical approaches have been proposed to study deformations in graphene 
\cite{Neto09,Winkler,Linnik,Salvador13,Kerner,Salvador14a,Salvador14b}. 
The most common one is a combination of the Tight-Binding Hamiltonian (TB) and linear elasticity 
to derive a Dirac effective equation \cite{Neto09}. 
Under such approach, pseudomagnetic fields appear, although lattice deformations were not
included in the original derivation \cite{Kitt}.
In the case of strain, recent works have included these considerations starting from different 
treatments \cite{SalvadorSSC,Salvador13,Kitt,Kitte13}. Still, there are some problems with such approach 
\cite{Oliva13} because a common confusion is the assumption 
that the Dirac cone tips $\bm{K}_D$ in the new deformed lattice coincides with the strained high symmetry 
points $\bm{K}$ and $\bm{K'}$. 

In the present work, we calculate analitically the group velocity surfaces for the important case of uniform strain, 
which can be solved without the  usual perturbative analysis of the Dirac equation. For this goal, the TB
approximation has been used. The Fermi velocities are obtained by looking at the appropiate points 
in the reciprocal space. Thus, our results are more general and include the Dirac theory of strain as a limiting case. 
Additionally, we found that even for a realistic value of pure shear strain, a mixed Dirac-Schr\"{o}edinger behavior can arise, 
suggesting that the Dirac theory has to be modified. In fact,
this behavior has been obtained in other cases \cite{Pereira09a,Pedro15}.  

The layout of this paper is the following. In Sec. 2, we describe electron behavior in graphene 
under uniform strain. Then a dispersion relation is obtained using TB aproximations. For 
this relation, we display the surfaces and contour plots for the particular 
case of pure shear strain.  In Sec. 3, we derive the group velocity for electrons, and analyze the 
pure uniaxial and pure shear strain cases. Finally, in Sec. 4, we give our conclusions.

\section{Electrons in Strained Graphene}
\begin{figure}
\center
\includegraphics[width =8cm,height=8cm]{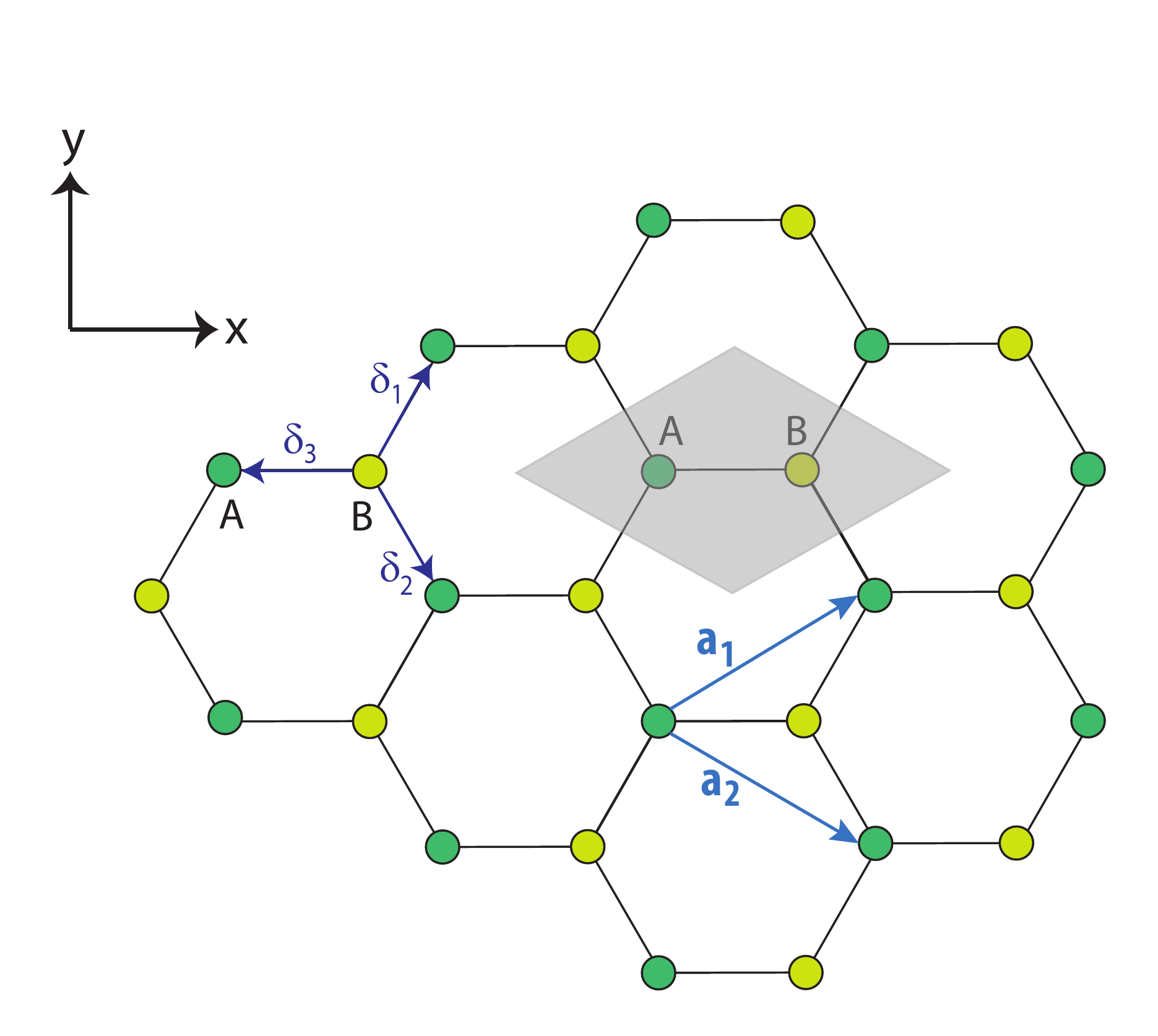}
\leavevmode \caption {(Color online) Graphene lattice and the sublattices A and B.
The associated unitary and first neighbour vectors are also shown.}
\label{RedGrafeno}
\end{figure}

Graphene is formed by a single atomic layer of carbon atoms arranged in an hexagonal estructure, as shown 
in Fig. \ref{RedGrafeno}. The structure can be described in terms of two triangular sub-lattices, A and B, 
with a basis of two atoms per unit cell. The lattice unit vectors are given by 
$\bm{a}_1$ and $\bm{a}_2$ and the three nearest-neighbor 
vectors can be written as $\bm{\delta}_1$, $\bm{\delta}_2$ 
and $\bm{\delta}_3$ \cite{Neto09}. Likewise, there are two reciprocal-lattice vectors given by 
$\bm{b}_1$ and $\bm{b}_2$, and two inequivalent special points at the corners of the graphene BZ, 
called high symmetry points $\bm{K}_0$ and $\bm{K'}_0$ \cite{Neto09}. 
For unstrained graphene, the tips of the Dirac cones (or the Dirac points $\bm{K}_D$) 
are located at the $\bm{K}_0$ and $\bm{K'}_0$ points. 

In the case of a uniform strain, if the vector $\bm{r}$ represents the positions of the carbon atoms 
in the undeformed graphene, its deformed counterpart is given by 
$\bm{r}^{\prime}=(\bm{I}+\bm{\epsilon})\cdot\bm{r}$, where $\bm{I}$ 
is the $2\times2$ identity matrix and $\bm{\epsilon}$ is the uniform strain 
tensor. The lattice unit and nearest-neighbor vectors are thus as follows, 
$\bm{a}_i^{\prime}=(\bm{I}+\bm{\epsilon})\cdot\bm{a}_i \quad (i=1,2)$ and 
$\bm{\delta}_n^{\prime}=(\bm{I}+\bm{\epsilon})\cdot\bm{\delta}_n\quad(n=1,2,3)$, 
while the reciprocal-lattice vectors are deformed as, 
$\bm{b}_i^{\prime}=(\bm{I}+\bm{\epsilon})^{-1}\cdot\bm{b}_i \quad (i=1,2)$. The new high symmetry points in the corners of the first BZ of the strained 
reciprocal lattice are obtained by construction of the Wigner-Seitz primitive cell and 
can written in general as
\begin{align}
 \bm{K}=\bm{M}^{-1}_{1}\bm{C}_1 \quad \text{and} \quad \bm{K}^\prime=\bm{M}^{-1}_{2}\bm{C}_2
 \label{K}
\end{align}
with
\begin{align*}
 \bm{M}_{i}=\left(\begin{array}{cc} 
(b_i^\prime)_x & (b_i^\prime)_y \\
(b_1^\prime)_x+(b_2^\prime)_x & (b_1^\prime)_y+(b_2^\prime)_y
\end{array} \right)
\end{align*}
and 
\begin{align*}
 \bm{C}_i=\frac{1}{2}\left(\begin{array}{c} 
\|\bm{b}_i^\prime\|  \\
\|\bm{b}_1^\prime+\bm{b}_2^\prime\| 
\end{array} \right),
\end{align*}
where $(b_i^\prime)_x$ and $(b_i^\prime)_y$ are the $x$ and $y$ components of the deformed reciprocal vectors  $\bm{b}_i^{\prime}$ ($i=1,2$).
\begin{figure*}
\center
\includegraphics[width =14cm,height=7cm]{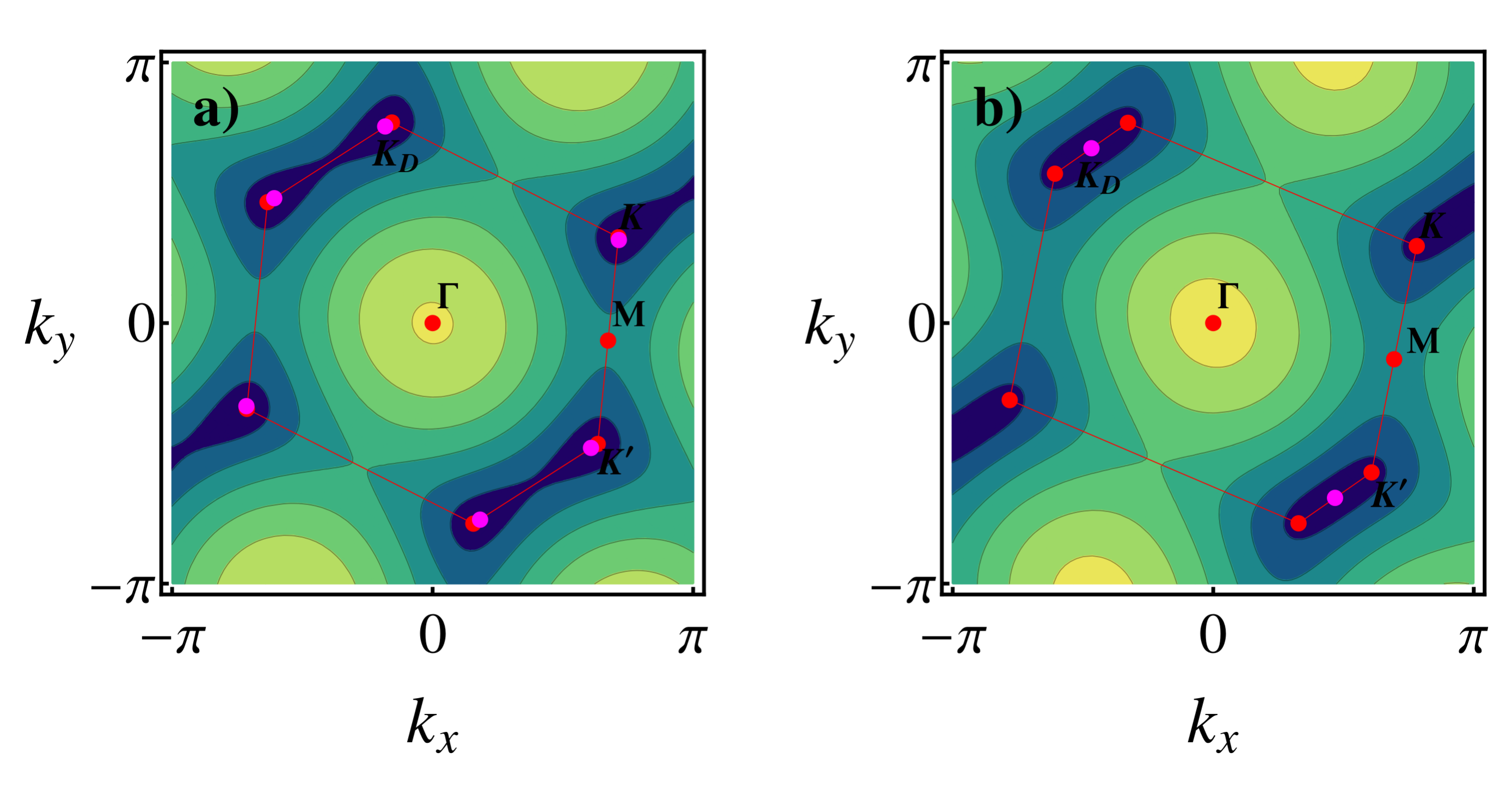}
\leavevmode \caption {(Color online) Contour plot for the energy for a shear strain given by $\epsilon_{xx}=\epsilon_{yy}=0$,  and a) $\epsilon_{xy}=0.1$ 
and b) $\epsilon_{xy}=0.2$. The first Brioullin zone of the strained reciprocal lattice is presented with red lines.
The high symmetry points $\bm{K}$ and $\bm{K^\prime}$ (Eq. \ref{K}) are indicated with red circles. The pink circles correspond to the position of
the Dirac cones $\bm{K}_D$, where the Fermi energy is located. Notice how two Dirac cones merges into one in case b) and 
do not have the same position as $\bm{K}$ and $\bm{K^\prime}$.}
\label{contour1}
\end{figure*}

To obtain the electronic properties of graphene under uniform strain, we use the nearest-neighbor TB Hamiltonian \cite{Oliva13}
\begin{equation}
\bm{H}=-\sum_{\bm{r}^{\prime},n} t_{n}^{\prime}\bm{a}_{\bm{r}^{\prime}}^{\dag} \bm{b}_{\bm{r}^{\prime} +\bm{\delta}_{n}^{\prime}}+\text{H.c.}, 
\end{equation}
where $\bm{r}^{\prime}$ runs over all sites of the deformed honeycomb lattice and the hopping integral $t_{n}^{\prime}$ 
varies due to the modification in the carbon-carbon distances as $t_{n}^{\prime}=t\exp[-\beta(|\bm{\delta}_{n}^{\prime}|/a - 1)]$ \cite{Ribeiro}, 
with $\beta\approx3$ and $t\approx2.7$ eV is the hopping energy for unstrained graphene \cite{Neto09}.  The operators 
$\bm{a}_{\bm{r}^{\prime}}^{\dag}$ and $\bm{b}_{\bm{r}^{\prime}+\bm{\delta}_{n}^{\prime}}$ correspond to creating and 
annihilating electrons on the $A$ sublattice position $\bm{r}^{\prime}$ and $B$ sublattice position $\bm{r}^{\prime}+\bm{\delta}_{n}^{\prime}$, 
respectively. Now, using the Fourier representation for these operators, the  previous Hamiltonian can 
be written as \cite{Oliva13}
\begin{equation}
\bm{H}=-\sum_{\bm{k},n} t_{n}^{\prime} e^{-i\bm{k}\cdot(\bar{\bm{I}} +
\bar{\bm{\epsilon}})\cdot\bm{\delta}_{n}} \bm{a}_{\bm{k}}^{\dag}
\bm{b}_{\bm{k}} + \text{H.c.},
\label{k-H}
\end{equation}
which finally leads to the closed dispersion relation for graphene under  uniform strain \cite{Oliva13}
\begin{equation}\label{GDR}
E(\bm{k})= \pm\left|\sum_{n} t_{n}^{\prime}
e^{-i\bm{k}\cdot(\bar{\bm{I}} +
\bar{\bm{\epsilon}})\cdot\bm{\delta}_{n}}\right|.
\end{equation}
It can be stressed that in our work, the latest equation has been developed more 
explicitly, which leads to
\begin{equation}
 E(\bm{k})=\pm\sqrt{\gamma+g(\bm{k})},
  \label{Ekxky}
\end{equation}
where
\begin{align*}
 g(\bm{k})=\sum_{n=1}^3\sum_{s>n}^3 t_n^{\prime} t_ s^{\prime}\cos\left[\bm{k}\cdot(\bm{I}+\bm{\epsilon})\cdot(\bm{\delta}_n-\bm{\delta}_s)\right]
\end{align*}
and 
\begin{equation*}
 \gamma=\sum_{n=1}^{3}{t_n^{\prime}}^2.
\end{equation*}
Let us now explore the strain effects on the energy dispersion relation. 
As it was previously explained in the introduction, when an uniform strain is applied
the reciprocal lattice is also
strained. Thus, the first BZ is modified, 
i.e., its original hexagonal form is varied to a polygonal form, as shown 
for the  particular cases of pure shear strain along the armchair 
direction: $\epsilon_{xx}=\epsilon_{yy}=0$, $\epsilon_{xy}=\epsilon_{yx}=0.1$ 
(see red lines Fig. \ref{contour1} a)) and $\epsilon_{xy}=\epsilon_{yx}=0.2$  
(see red lines Fig. \ref{contour1} b)). 
In the same figure \ref{contour1}, along with the first BZ, we present the contour plot of the energy obtained from
Eq. (\ref{Ekxky}). 

Once the first BZ and the energy surfaces are obtained, we need to locate the 
Dirac points $\bm{K_D}$ using the condition $E(\bm{K_D})=E_{F}$, which corresponds 
to electrons at the Fermi level. 
By applying  this condition, the Dirac points are indicated as pink circles in   Fig. \ref{contour1}.
The most important conclusion from the figure is that such points are no longer located at the 
high symmetry points $\bm{K}$ and $\bm{K^\prime}$ (red circles) of the corners of the first BZ (Eq. \ref{K}), since they are shifted to the saddle point.  
This plot ilustrates an issue that has been disregarded in the literature concerning graphene.

As deformation increases (up to 20$\%$), the Dirac points merge into the saddle point
and a gap opens, which is consistent with the results 
in references \cite{Pereira09a,Montambaux}. Furthermore, in this 
critical point, the dispersion relation is linear along 
one direction ((relativistic Dirac behavior)
and quadratic along the other one (nonrelativistic Schr\"{o}dinger behavior). Therefore, 
the Dirac theory needs to be modified.

Summarizing the above, the effects caused in graphene under uniform strain are the 
following: 
\begin{itemize}
\item i) The Dirac points are shifted from the corners of the strained BZ. 
\item ii) The Dirac equation is no 
longer suitable for long strain ($\geq 20\%$), since for particular cases a
Dirac-Schr\"{o}dinger behavior is observed and furthermore one might expect
significant nonlinear corrections.  It follows that the anisotropic 
Fermi velocity is not longer valid in these regimens. Therefore, we must consider 
a more general velocity to understand the electron behavior. This is done through the 
calculation of the group velocity,  as we will discuss in the following section.
\end{itemize}

\section{Group velocity}
\begin{figure*}
\center
\includegraphics[width =14cm,height=7cm]{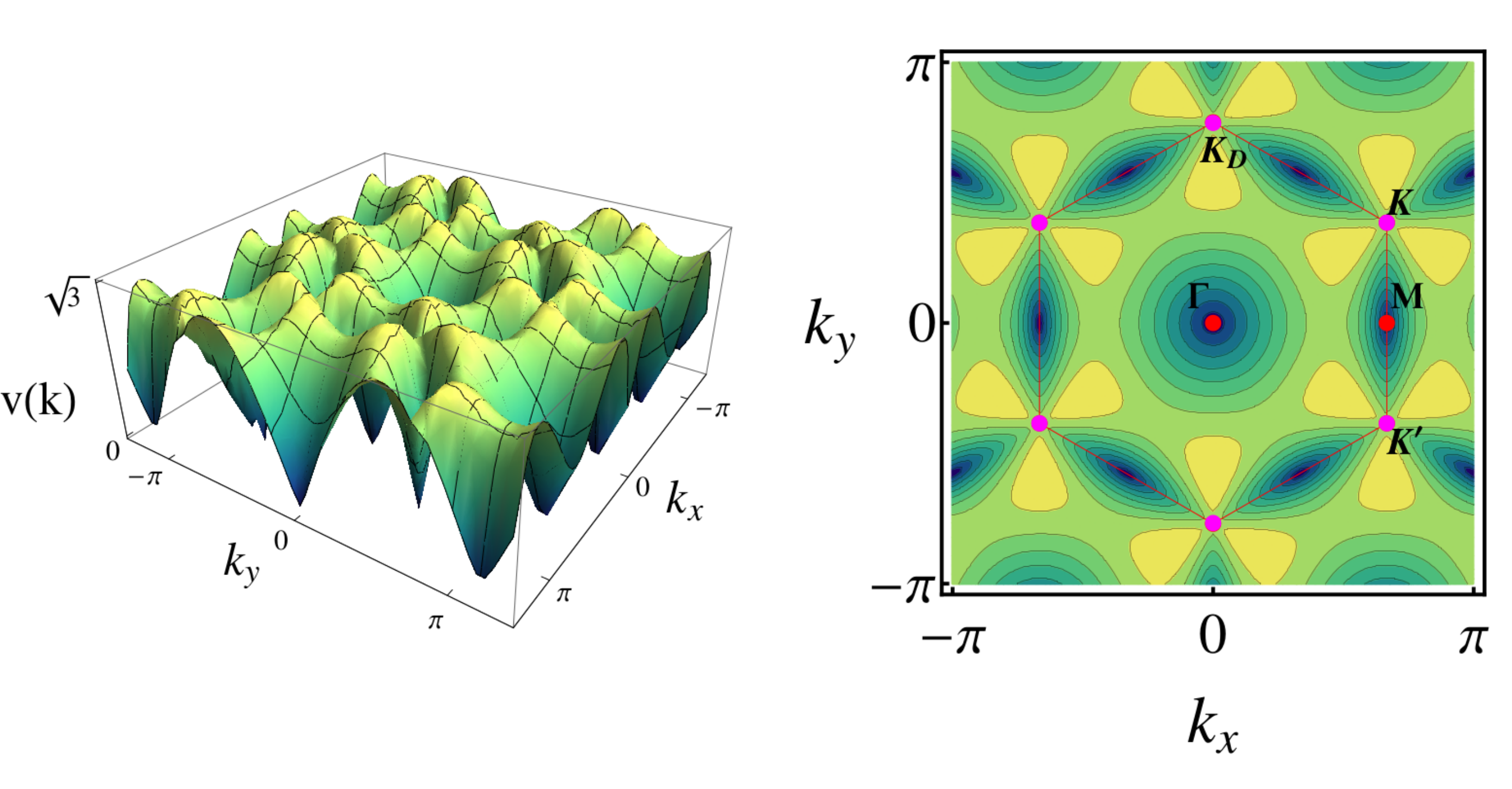}
\leavevmode \caption {(Color online) Surface and contour plot of the group velocity norm for 
pure graphene.}
\label{energy}
\end{figure*}

\begin{figure*}
\center
\includegraphics[width =14cm,height=7cm]{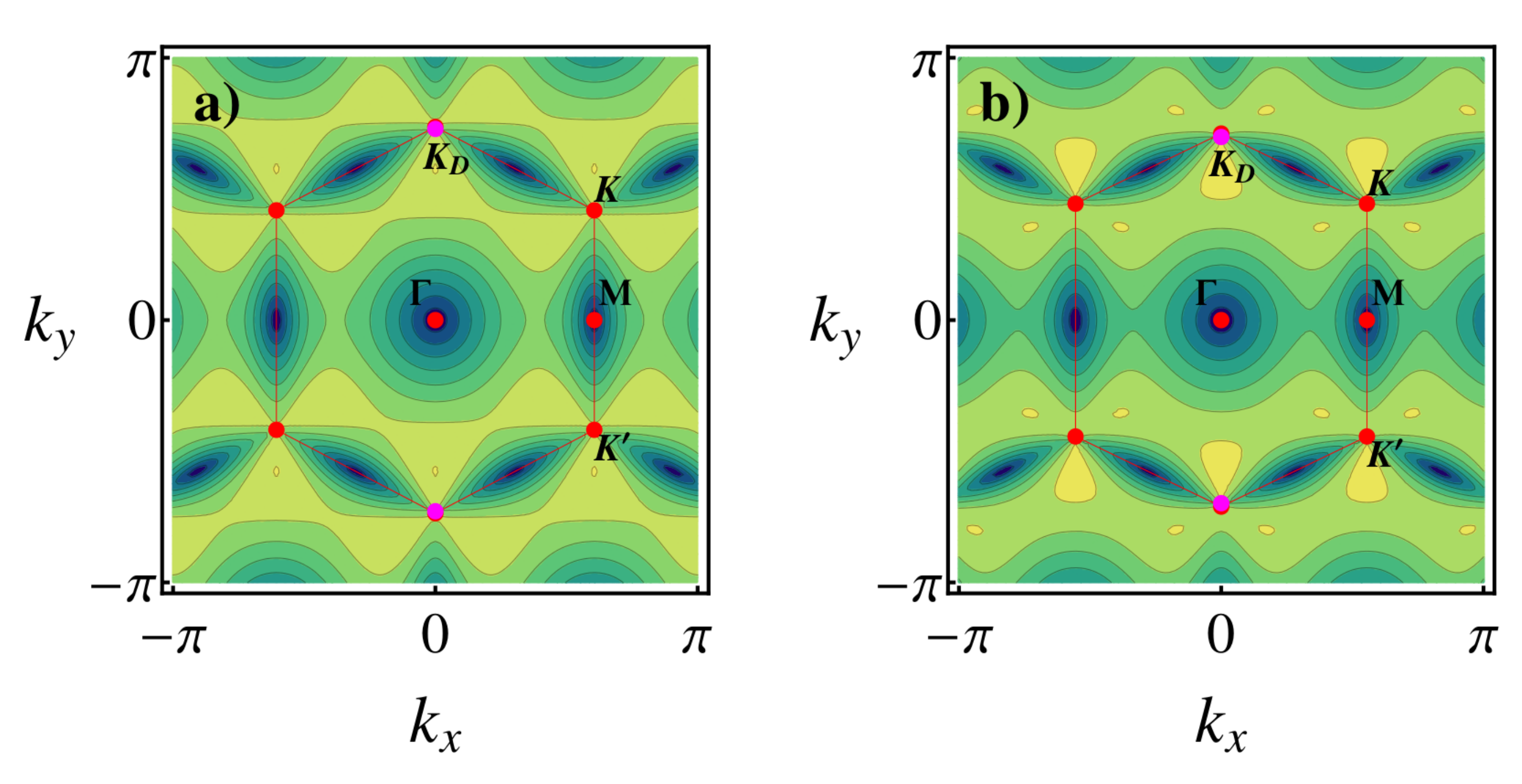}
\leavevmode \caption {(Color online) Countour plot of the group velocity for a uniaxial uniform strain
given by $\epsilon_{xy}=\epsilon_{yy}=0$, and with a) $\epsilon_{xx}=0.1$ and b) $\epsilon_{xx}=0.2$.
The first Brioullin zone of the strained reciprocal lattice is presented with red lines.
The high symmetry points $\bm{K}$ and $\bm{K^\prime}$ (Eq. \ref{K}) are indicated with red circles. The pink circles correspond to the position of
the Dirac cones $\bm{K}_D$, where the Fermi energy is located. Notice how $\bm{K}_D$
do not have the same position as $\bm{K}$ and $\bm{K'}$.}
\label{exx}
\end{figure*}

In the literature, the basic properties of electron transport phenomena
in a crystal are described in terms of Bloch waves with
wave vectors $\bm{k}$ \cite{Mizutani}. Using these waves, we can 
build a dispersive wave packet with a certain group velocity. 
It can be shown quite generally that the mean electron velocity 
is given by the group velocity of the wave packet \cite{Mizutani}
\begin{equation}
 \bm{v}(\bm{k})=\nabla_{\bm{k}}E(\bm{k}),
 \label{vxvy}
\end{equation}
where $\nabla_{\bm{k}}$ is the gradient operator in $\bm{k}$-space. From this equation, the real-space motion of the electron can be described. 
Here we are  interested in the behavior of electrons in graphene under uniform strain. 
Thus, by substituting Eq. (\ref{Ekxky}) into Eq. (\ref{vxvy}) we obtain
\begin{equation}
 \bm{v}(\bm{k})= \pm \frac{1}{2E(\bm{k})}\nabla_{\bm{k}}g(\bm{k}).
 \label{vectorv}
\end{equation}
The components $x$ and $y$  of  $\bm{v}(\bm{k})$ are given by
\begin{align}
&v_l(\bm{k})=\pm\sum_{n=1}^3\sum_{s>n}^3[(1+\epsilon_{ll})(\delta_n^l-\delta_s^l)+\epsilon_{lm}(\delta_n^m-\delta_s^m)]\nonumber\\
&\qquad\times t_n^{\prime}t_s^{\prime}\frac{\sin[\bm{k}\cdot(\bm{I}+\bm{\epsilon})\cdot(\bm{\delta}_n-\bm{\delta}_s)]}{\sqrt{\gamma+g(\bm{k})}},
\end{align}
where $l,m=\{x,y\}$ and $l\ne m$. The group velocity norm is given by $v(\bm{k})=|\bm{v}(\bm{k})|$.

In the Fig. \ref{energy} we plot the surfaces and contour of the group velocity
norm $v(\bm{k})$ for pure graphene. It is important to note that 
at low energies and in the vicinity of the Dirac points, $v(\bm{k})$ is isotropic, 
coinciding with the Fermi velocity. However, as we move away from the Dirac point 
(corresponding to nonlinear  corrections to the Dirac cone) a trigonal warping appears, 
giving rise to an anisotropic behavior 
(see contour Fig. \ref{energy}). Furthermore, it can be observed that in the directions 
where the trigonal warping appear, $v(\bm{k})$ increases, 
while in other directions decreases drastically. These results do not appear when 
the Dirac theory is used. Therefore, if we want a 
more complete understanding of the behavior of electrons in the
energy bands, nonlinear corrections and directions should be considered, since
strain effects enhance such features, as we discuss below. 

\begin{figure*}
\center
\includegraphics[width =14cm,height=7cm]{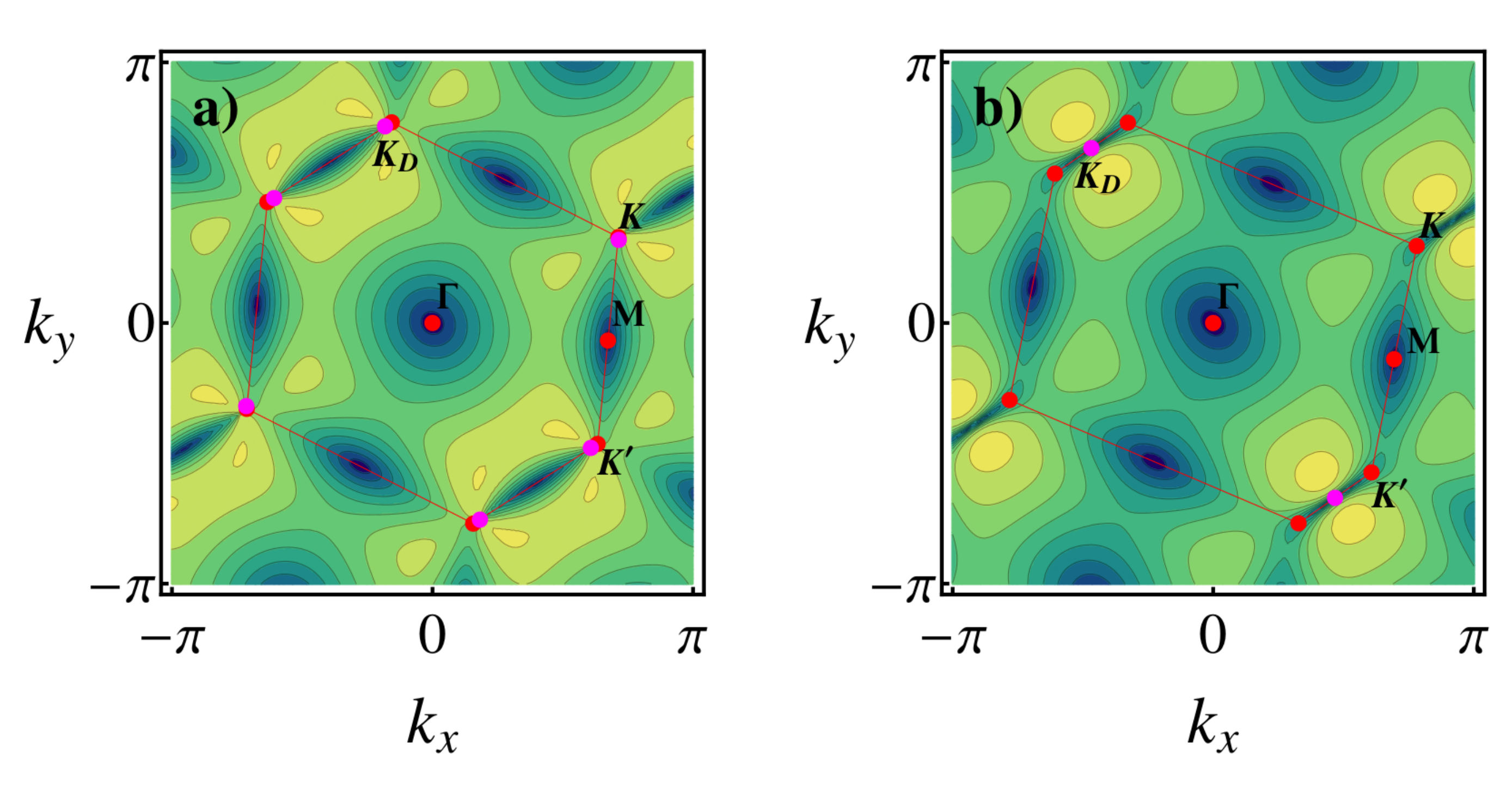}
\leavevmode \caption {(Color online) Countour plot of the group velocity for a shear strain
given by $\epsilon_{xx}=0$, $\epsilon_{yy}=0$, and with a) $\epsilon_{xy}=0.1$ and b) $\epsilon_{xy}=0.2$.
The first Brioullin zone of the strained reciprocal lattice is presented with red lines.
The high symmetry points $\bm{K}$ and $\bm{K^\prime}$ (Eq. \ref{K}) are indicated with red circles. The pink circles correspond to the position of
the Dirac cones $\bm{K}_D$, where the Fermi energy is located. Notice
how two Dirac cones merge in case b) and do not have the same position as $\bm{K}$ and $\bm{K^\prime}$. From the countor plot, is clear
that the Fermi velocity is constant (Dirac behavior) in the principal axis of the the shear,
while it follows a parabolic behavior (Schroedinger) in the perpendicular direction.}
\label{exy}
\end{figure*}

We analyze the particular cases of pure uniaxial and pure shear strain with 
$\epsilon_{xy}=\epsilon_{yy}=0$, $\epsilon_{xx}=0.1,0.2$  and 
$\epsilon_{xx}=\epsilon_{yy}=0$, $\epsilon_{xy}=0.1,0.2$, 
as shown in Fig. \ref{exx} and Fig. \ref{exy}, respectively. In these Fig. \ref{exx} and Fig. \ref{exy}, the contours plots of the velocity norm $v(\bm{k})$ are presented.
Overimposed to these contours, we present the first BZ of the strained reciprocal lattice with red lines. 
Likewise, the high symmetry points (Eq. \ref{K}) are indicated with red circles and the pink circles correspond to the position of
the Dirac point $\bm{K}_D$ where the Fermi energy is located. Notice again how the $\bm{K}_D$ points
do not have the same position as the high symmetry points $\bm{K}$ and $\bm{K'}$. The effect for shear strain 
is much more pronounced. 

On the other hand, the effects caused by the deformation in the velocity surfaces are the following: 
\begin{itemize}
\item  i) As seen in \ref{exx} and Fig. \ref{exy}, the Fermi velocity is no longer isotropic. 
Instead, it becomes strongly anisotropical.
\item ii) The surfaces do not display the trigonal symmetry, and instead, they present the symmetry of the
    corresponding strained reciprocal lattice.
\item iii) The trigonal warping observed in pure graphene disappears, as in (Fig. \ref{exx} a)).  
It reappears in (Fig. \ref{exx} b)) but with deformed angles which follow the symmetry of the strained reciprocal lattice.
This new warping is strongly modulated, and it touches the Dirac points (Fig. \ref{exx} b)). Since the 
warping is associated with non-linear terms, this suggests that non-linearity is important in order to describe such cases. As
a result, a pure Dirac equation is not longer valid. For the shear strain, in Fig. \ref{exy} a) the trigonal warping is deformed,
until in Fig. \ref{exy} b) becomes two lobules reflecting  the symmetry of the strained reciprocal space. 
\item iv) When the Dirac cones merges by shear strain as in Fig. \ref{exy} b), the group velocity is zero
along one of the pricipal axis of the deformation. This is just the consequence of the energy having a 
parabolic (Schr\"{o}dinger) behavior with a gap opening \cite{Pereira09a}. 
\end{itemize}

\section{Conclusions}

We have obtained the electron's group velocity for graphene under uniform strain using 
the tight-binding approximation. Our results indicate that the velocity is strongly
anisotropic and that the trigonal warping is deformed to follow the symmetry of the
strained reciprocal lattice. As strain increases, this warping touches the
Dirac points. Thus, we found that non-linearity is very important in order to 
describe electrons in a proper way near the Fermi energy for strain, since the trigonal warping 
observed in graphene touches the Dirac point and gets modulated by the symmetry
of the strained reciprocal lattice. As a result, a Dirac equation kind of
approximation is not  longer valid for such cases.

Finally, our closed analytical expressions for the electron velocities are useful in order to 
calculate electronic, thermal and optical properties of strained graphene.  

\section*{Acknowledgments}

G. Naumis thanks the program DGAPA-PASPA for a sabatical shoolarship. W. G\'omez-Arias thanks CONACyT for a master schoolarship. 
This work was funded by DGAPA-PAPIIT proyect 102513. Calaculations were performed at DGTIC-UNAM supercomputer center. 

\bibliography{mybibfile}

\end{document}